\documentclass{styles/optica-article}

\journal{opticajournal} 

\articletype{Research Article}

\usepackage{float}
\usepackage{siunitx}
\usepackage{accents}
\usepackage{pdfpages}

\usepackage{fixme}
\fxsetup{status=draft} 

\begin{document}

\title{All-Optical Field-Resolved Spectroscopy With Interferometric Nonlinear Cross-Correlations}

\author{Felix Ritzkowsky,\authormark{1,2,*} Gian Luca Dolso,\authormark{1} Benjamin M. Mazur,\authormark{1} Matthew Yeung,\authormark{1} and Phillip D. Keathley \authormark{1,*}}

\address{\authormark{1}Massachusetts Institute of Technology, 77 Massachusetts Ave., 02139 Cambridge MA.\\

\authormark{2}Deutsches Elektronen Synchrotron (DESY) and Center for Free-Electron Laserscience (CFEL),Notkestr. 85, 22607, Hamburg, Germany.\\}

\email{\authormark{*}pdkeat2@mit.edu, felix.ritzkowsky@desy.de} 


\begin{abstract*} 
Direct time-domain measurements of electric fields enable sub-cycle spectroscopy of light–matter interactions, but established techniques such as electro-optic sampling are constrained in their bandwidth by gate-pulse duration and phase-matching limitations\cite{srivastavaNearpetahertzFieldoscopyLiquid2024}. Alternative approaches have emerged in recent years based on asymmetric interferometric nonlinear cross-correlations with highly nonlinear media \cite{parkDirectSamplingLight2018,biontaOnchipSamplingOptical2021,yeungLightwaveelectronicHarmonicFrequency2024,liuAllopticalSamplingFewcycle2021,liuSingleshotMeasurementFewcycle2022}, and have demonstrated, for example, the field-resolved study of exciton ensembles \cite{yeungFieldresolvedObservationExciton2026}. However, these nonlinear cross-correlation-based techniques have been benchmarked almost exclusively by self-referenced pulse characterization rather than by their quantitative spectroscopic performance, and all-optical approaches have received less attention than those based on direct charge emission. Here we extend all-optical asymmetric interferometric cross-correlation to higher nonlinearities in sub-wavelength films and demonstrate field-resolved spectroscopy of the free-induction decay of two ro-vibrational bands of ambient water vapor with a performance comparable to state of the art electro-optic sampling.
The measurement spans \SI{190}{\tera\hertz} of bandwidth (\SI{80}{\tera\hertz} to \SI{270}{\tera\hertz}) with sub-\SI{500}{\giga\hertz} spectral resolution, a spectral intensity dynamic range of six orders of magnitude, and a field-strength noise floor of \SI{100}{\kilo\volt\per\meter}. We anticipate the rapid adoption of here presented all-optical sampling to many experimental settings and a broad impact beyond the ultrafast optics research community as it is drastically simplified in comparison to ionization based techniques and allows the translation of electro-optic-sampling-level sensitivity into higher frequency ranges not previously accessible by conventional tools.

\end{abstract*}
\section{Introduction}
The electric field of a light wave directly governs how light interacts with the electronic response of matter. It can drive processes that depend mainly on the pulse envelope and are often approximately linear, such as exciton dynamics \cite{yeungFieldresolvedObservationExciton2026,wilsonExcitonsEmergentQuantum2021} and molecular vibrational responses \cite{pupezaFieldresolvedInfraredSpectroscopy2020}, or it can also induce fully field-driven, sub-cycle, and strongly nonperturbative phenomena, including Landau-Zener-Stückelberg electron interference in graphene \cite{higuchiLightfielddrivenCurrentsGraphene2017,boolakeeLightfieldControlReal2022}, attosecond charge transport in nanosystems \cite{rybkaSubcycleOpticalPhase2016,putnamOpticalfieldcontrolledPhotoemissionPlasmonic2017}, and lightwave-driven quasiparticle collisions in semiconductors \cite{langerLightwavedrivenQuasiparticleCollisions2016}. However, conventional methods of studying these phenomena are generally indirect, relying on observing detected currents\cite{rybkaSubcycleOpticalPhase2016,higuchiLightfielddrivenCurrentsGraphene2017,boolakeeLightfieldControlReal2022} or measuring the spectral amplitudes of scattered fields. Critically, these approaches do not interrogate the phase of the electrical field and thus cannot directly access the temporal dynamics of light-matter interaction.

For decades, electro-optic sampling (EOS) has been the dominant tool in the infrared and terahertz spectral domains, with spectral coverage spanning gigahertz \cite{valdmanisPicosecondElectroOptic1982,wuFreespaceElectroopticSampling1995,sellFieldresolvedDetectionPhaselocked2008} to short-wave infrared (SWIR) \cite{srivastavaNearpetahertzFieldoscopyLiquid2024} frequencies, and dynamic range sufficient to resolve quantum vacuum fluctuations \cite{riekDirectSamplingElectricfield2015}. However, EOS is fundamentally limited by the gate pulse duration and phase-matching constraints in the nonlinear medium, requiring high average power laser systems to reach the SWIR regime \cite{keiberElectroopticSamplingNearinfrared2016,srivastavaNearpetahertzFieldoscopyLiquid2024}. Recent extensions of EOS to higher order nonlinearities have demonstrated detection at high frequencies, but have not yet shown signal fidelity beyond standard pulse characterization, which does not require high dynamic range or measurement bandwidth beyond the pulse spectrum \cite{ridenteElectroopticCharacterizationSynthesized2022,ziminUltrabroadbandAllopticalSampling2022}.

A simple and elegant approach for field-resolved spectroscopy was hiding in plain sight. Interferometric autocorrelations (IAC) have been a cornerstone in the characterization of femtosecond optical pulses since its first demonstration by Diels et al. in 1985 \cite{dielsControlMeasurementUltrashort1985} and the introduction of iterative retrieval for the full phase characterization by Naganuma et al. \cite{naganumaGeneralMethodUltrashort1989}. However, it took over 30 years to realize that the phase ambiguity of IAC can be completely lifted by unbalancing the intensities in the interferometer and thus can be operated in a way that directly measures the electric field of the pulse under test. 

The first demonstration of IAC in this novel regime was performed by Park et al. \cite{parkDirectSamplingLight2018}, showing that the effective perturbation of highly nonlinear tunneling ionization in air by the weak arm of the interferometer produces an ionization signal proportional to the instantaneous electric field of the weak signal. This scheme has been subsequently demonstrated in several other highly nonlinear systems, including ionization from nanostructured systems \cite{biontaOnchipSamplingOptical2021,yeungLightwaveelectronicHarmonicFrequency2024,yeungBandwidthLightwaveDrivenElectronic2025}, multiphoton absorption in semiconductors \cite{liuSingleshotMeasurementFewcycle2022} and nonlinear fluorescence in ZnO \cite{truongScanlessLaserWaveform2025}. Furthermore, a recent experiment by Gliserin et al. has shown that it is also sufficient to simply perturb the second-harmonic generation in a $\chi^{(2)}$-medium \cite{gliserinCompleteCharacterizationUltrafast2022}, rendering it conceptually identical to the original IAC as proposed by Diels et al. However, no one has demonstrated field resolved spectroscopy with a signal fidelity comparable to EOS. Most demonstrations have been exclusively limited to self-referenced pulse characterization, limited also by the amplitude-weighted nature of the iterative retrieval process, yielding an SNR incompatible with spectroscopic applications\cite{choReconstructionAlgorithmTunneling2021,wieseUniversalWaveformresolvingDual2024,leungUniversalInlineWaveform2026}.

\begin{figure}[b]
    \centering
    \includegraphics[width=16cm]{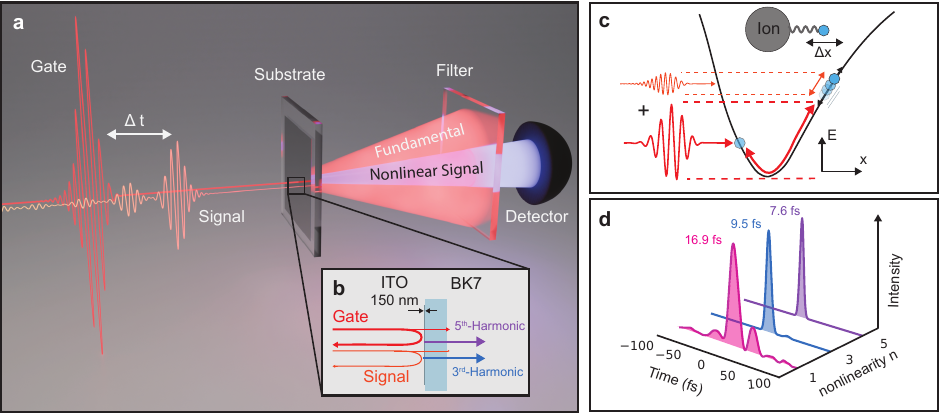}
    \caption{\textbf{Interferometric Nonlinear Cross-Correlation Concept: a}, depiction of the sampling process with a strong gate pulse and a weak signal pulse, collinearly interfering on the ITO thin film. The strong gate drives nonlinear emission up to the 5th harmonic, which is perturbed by the weak signal field. The nonlinear signal is separated by a shortpass filter and detected with a silicon photodetector. \textbf{b}, depiction of the nonlinear signal generation in the thin film ITO system. The ITO has a thickness of \SI{150}{\nano\meter} on a BK7 substrate. The gate and signal fields are partially reflected at the ITO interface, while the third and fifth harmonics transmit through. \textbf{c}, atomic potential of the electron bound to its parent ion. The electron is driven in the anharmonic potential out of equilibrium by the strong gate field creating a nonlinear response of the system. This nonlinearity is perturbed by the weak signal pulse, imprinting its electric field on the perturbed nonlinearity. \textbf{d}, effective temporal response of the gate pulse based on the system nonlinearity. Higher nonlinearities effectively reduce the temporal gate duration and thus extend the accessible bandwidth.}
    \label{fig:overview}
\end{figure}
In this work we demonstrate that all-optical nonlinear interferometric cross-correlations (ICC)~\cite{gliserinCompleteCharacterizationUltrafast2022} can achieve electric field detection comparable in resolution and dynamic range to EOS in the SWIR, dramatically reducing necessary experimental complexity. In addition to quantitative benchmarking of the technique, we show the detection of free-induction decay (FID) signals from two ro-vibrational bands of ambient water vapor, quantitatively matching absorption data provided by the HITRAN database \cite{gordonHITRAN2020MolecularSpectroscopic2022}. Our approach yielded a direct measurement of these optical fields down to \SI{100}{\kilo\volt\per\meter} over several picoseconds with sub-femtosecond resolution. 

With access to the full complex field, we performed time-frequency analysis of the free-induction decay waveforms that reveals temporal structure and vibrational mode beating that would be obscured in conventional spectroscopy methods. Through this analysis we show that we achieved a spectral resolution of sub-\SI{500}{\giga\hertz} with a spectral intensity dynamic range greater than 6 orders of magnitude, at a spectral bandwidth of \SI{190}{\tera\hertz} covering \SI{80}{\tera\hertz} to \SI{270}{\tera\hertz}.  This performance was enabled by extending the all-optical ICC approach to higher nonlinear orders (up to the 5th harmonic) as well as the use of sub-wavelength thin-films, here indium tin oxide (ITO), for harmonic generation to avoid phase-matching  constraints from conventional nonlinear crystals. Moreover, we demonstrate how performance metrics developed by the electrical engineering community for microwave mixers can be extended generally to such nonlinear cross-correlation techniques for objectively benchmarking performance. In particular we analyze the spurious free dynamic range (SFDR) of our measurements and use it to clarify that our measurements were operated in a regime that is quantitatively meaningful without necessitating iterative retrieval algorithms. 

\section{Sampling Framework}

Our methodology is based on the perturbation of the nonlinear emission from a solid medium, as laid out by Gliserin et al. \cite{gliserinCompleteCharacterizationUltrafast2022} and depicted in Fig. \ref{fig:overview} a. The gate field $E_G(t)$, a strong pulse, drives a nonlinear polarization response $P^{\left(n\right)}(t,\tau)$ to arbitrary order $n$, which is interferometrically perturbed by a factor of $s$ weaker signal field $E_S(t-\tau)$. In the perturbative regime considered here, $s \ll 1$. The nonlinear interferometric cross-correlation signal $I_\text{CC}(\tau)$ is detected and integrated on a slow photo-detector. To within a constant factor,

\begin{align}
        I^{(n)}_\text{CC}(\tau)
        \propto \int_{-\infty}^{\infty}
        \bigl\lvert P_n\left(t,\tau\right) \bigr\rvert ^{\,2} dt
        \propto \int_{-\infty}^{\infty}  \,  \Bigl\lvert\bigl(E_G(t)+sE_S(t-\tau)\bigr)^n  \Bigr\rvert^{\,2} dt.    
\end{align}
In the microscopic picture, this process can be understood as a perturbation of an electron's trajectory around its parent ion in an anharmonic potential, driven by a strong gate pulse (see Fig. \ref{fig:overview} c). Only for the brief periods where the electric field is driving the electron up the potential, the anharmonic potential contributions are relevant, therefore providing a periodic time-gate for the perturbation.  

Considering only the relevant terms describing the sampling process, assuming an instantaneous nonlinear response and ideal phase matching, we have that

    \begin{align}
    I_\text{CC}(\tau) \propto \int_{-\infty}^{\infty} \, \underbrace{\bigl\vert E_G^n(t)\bigr\vert^2}_\text{DC} + n\underbrace{E_G^*(t) \bigl\vert E_G^{n-1}(t)\bigr\vert^{2}}_{\text{Gate Kernel: } G(t)} \cdot \, sE_S(t-\tau) + \underbrace{D(t,\tau)}_{\text{Distortion}} dt .     
    \end{align}
Note that the full analytical derivation of the sampling equation can be found in the supplementary section 1 and in references \cite{choReconstructionAlgorithmTunneling2021,gliserinCompleteCharacterizationUltrafast2022,parkDirectSamplingLight2018}. The first term in the integrand results in a DC background that is independent of the signal and can be easily subtracted. The second term is the correlation signal that contains the information about the electric field of the signal pulse, and is linear in the signal field. The gate kernel $G(t) = E_G^*(t) \vert E_G(t)^{n-1}\vert^2$, which sets the temporal resolution of the measurement, transforms the signal field into the measured output signal. This correlation kernel is effectively a temporally gated version of the gate field $E_G(t)$, and it directly follows that with increasing nonlinearity this function becomes shorter in duration. This relationship is visualized in Fig. \ref{fig:overview} d, which shows the intensity envelope of the gate function (pink) as characterized by a dispersion scan (d-scan) method\cite{mirandaCharacterizationBroadbandFewcycle2012}. When calculating the gate functions for higher order nonlinearities of isotropic materials, $n=3$ (third-harmonic generation) in blue and $n=5$ (fifth harmonic generation) in purple, it is evident that with increasing nonlinear order not only does the gate become shorter, any spurious temporal features outside of the peak field are stripped away. In the limit $n\rightarrow\infty$, the gate function approaches a Dirac delta function, which produces an ideal,
artifact-free representation of the signal field $E_S(t)$. In particular,
the high nonlinearity removes long-lived pedestals that could otherwise
produce artifacts when interrogating faint spectroscopic signals in the
wings of the pulse, rendering the cross-correlation effectively
background-free in time.

Fig.~\ref{fig:transfer} shows the effective gate transfer functions for
$n=3$ and $n=5$. The 10-dB bandwidth broadens from \SI{150}{\tera\hertz}
at $n=3$ to approximately \SI{200}{\tera\hertz} at $n=5$, extending
detection beyond the spectral range of the gate field itself. Additionally, the spectral phase of $G(\omega)$ becomes flatter than that of the gate field $E_G(t)$. A flat phase of the gate kernel is critical, since any structure
in $\angle G(\omega)$ would be imprinted as spectral phase distortion on
the measured $E_S(\omega)$. Together with the temporal localization
established above, these properties allow direct readout of $E_S(t)$
from $I_\mathrm{CC}(\tau)$ without deconvolution or iterative
reconstruction.


The remaining term, $D(t,\tau)$, contains higher-order distortion components that are nonlinear in the signal field, neglecting terms beyond third order. The distortion term is analytically derived in equation Supplementary Information 1, and is analogous to distortions in RF mixer theory \cite{narayananTransistorDistortionAnalysis1967}. The mathematical framework developed in electrical engineering for describing the performance of microwave mixers provides us with a robust set of quantitative tools that we can leverage for objective benchmarking of IAC and ICC measurements.  In particular, we can use established metrics for quantifying nonlinear distortion and the reliable operating range without the need for numerical correction in post analysis.  

The distortion components are presented here and are separated by their respective frequency bands, and can be classified as harmonic or intermodulation distortion (IMD), 
\begin{align}
    D(t,\tau)= \underbrace{s^2 \binom{n}{1}^2 D_{0\omega}(t,\tau)}_\text{0$^\text{th}$ Harmonic Distortion}+\underbrace{s^2 \binom{n}{2} D_{2\omega}(t,\tau)}_{\text{2$^\text{nd}$ Harmonic Distortion}} + \underbrace{s^3 \binom{n}{2}\binom{n}{1}D_{\omega}(t,\tau)}_{\text{Intermodulation Distortion}} + \,\dots \quad.
\end{align}
The first term is causing a spurious signal components at the signal's 0th harmonic and $D_{0\omega}(t,\tau) \propto \vert E_S(t)\vert^2 $. The second term spectrally coincides with the 2nd harmonic of the signal with $D_{2\omega}(t,\tau)\propto E_S^2(t)$. In most cases, these harmonic distortion tones can be filtered from the signal spectrum. However, IMD, an important quantity in any nonlinear system \cite{narayananTransistorDistortionAnalysis1967}, yields mixing products that coincide with the signal spectrum, and thus cannot be trivially separated. Third-order IMD is proportional to $D_{\omega} \propto \vert E_S(t)\vert^2 E_S(t)$ and takes the same functional form as self-phase modulation. As $D_{\omega}$ scales nonlinearly with the signal amplitude, it is possible to define the spurious-free dynamic range (SFDR) of the measurement,
\begin{align}
\text{SFDR} = \frac{4}{s^4 \, n^2 (n-1)^2} ,    
\end{align}
see Supplementary Section 1 for more details. The SFDR defines the separation between the signal and the highest nonlinear distortion artifact. It is particularly important to define SFDR, as nonlinear distortions, in contrast to random noise, appear as a coherent signal, and could be mistaken for real measurement signals. 
The signal-to-gate amplitude ratio is chosen such that the distortion products are weaker than the smallest signal of interest, removing the need for algorithmic corrections. However, the SFDR is defined with respect to the strongest feature measured and follows it instantaneously, this also means that for weak trailing signatures the distortion will be substantially reduced to a locally smaller $s$, considerably relaxing the distortion constraint. For $n=5$, $\text{SFDR}\sim\frac{1}{100} s^{-4}$, corresponding to a \SI{60}{\decibel} SFDR for $s=0.01$. It is also useful to establish the 1-dB compression point $s_{1\text{dB}} = ({0.24(n-2)!}/{n!})^{1/2}$, which places an upper bound on $s$ to limit the inherent nonlinear deviation of the measured signal amplitude  to within \SI{1}{\decibel} or \SI{12}{\percent}.
 \newline
\section{Results and Discussion}
To demonstrate the all optical field sampling and the role of the nonlinearity, we performed an experiment using a dispersion balanced Mach-Zehnder interferometer and a thin-film ITO target as the nonlinear medium. A detailed description of the measurement setup is provided in the Methods section \ref{Methods:Setup} and illustrated in Extended Data Fig. \ref{fig:setup}.

The ITO film is a commercial \SI{150}{\nano\meter} film, sputter-coated on a \SI{3}{\milli\meter} BK7 substrate. The ITO fulfills three main purposes: first it allows for the controlled generation of a surface harmonic, due to its reflectivity for wavelengths above \SI{1.5}{\micro\meter}, causing only the film interface to generate nonlinear signal and not the bulk or the exit interface of the substrate. Second, the large $\chi^{(3)}$ gives comparatively larger THG signal, and third the thin film nature removes the phase matching constraint typically imposed by bulk media. The nonlinear emission, predominantly third- and fifth-harmonic emission, is separately filtered and detected by an amplified silicon photo-diode or spectrometer. 

To characterize the gate and the signal fields individually we measured the spectrum of the third-harmonic generation as a function of the ZnSe wedge insertion, realizing a surface third-harmonic d-scan in the identical nonlinear system also used for the field sampling experiments. We found slight deviations between the two pulses in spectral composition and pulse shape due to the different beam paths, with the gate measuring a duration of \SI{16.9}{\femto\second} and the signal of \SI{15.8}{\femto\second}, see Extended Data Fig. \ref{fig:dscan}.

\begin{figure}[h]
    \centering
    \includegraphics[width=16cm]{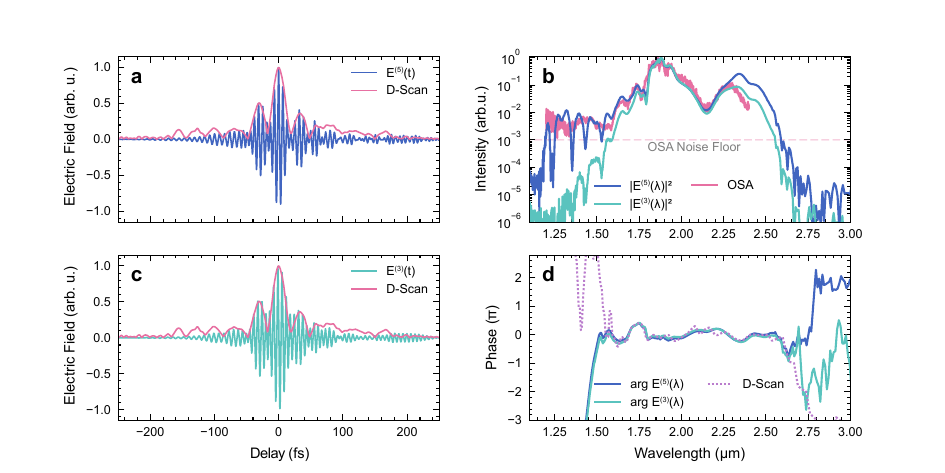}
    \caption{\textbf{Cross-Correlation Measurement Results in Time- and Spectral-Domain: a, c,} measured electric field waveforms as a function of time (delay). The electric field is proportional to the measured photodiode intensity of the nonlinear signal. The pink plot is showing the direct comparison to envelope shape as characterised by an in situ d-scan. \textbf{b,} spectral intensity of the electric fields from a, b for the two measurement cases, and SWIR optical spectrum analyzer. \textbf{d,} spectral phase from the two electric field measurements and the d-scan measurement. }
    \label{fig:data}
\end{figure}

The resulting fifth-harmonic ($E^{(5)} (t)$) and the third-harmonic ($E^{(3)} (t)$) ICC traces are shown in Fig. \ref{fig:data} a \& c. The electric field is overlapped with the measured d-scan envelope to highlight agreement and differences. For reference, we have also plotted the intensity envelopes of the results in comparison to the dispersion scan measurement in Extended Data Fig. \ref{fig:comparison}.  

We found differences in the measurement FWHM pulse durations. The d-scan method retrieved a duration of \SI{15.8}{\femto\second}, while the ICC based methods retrieved \SI{17.4}{\femto\second} for the third-harmonic case and \SI{13.5}{\femto\second} in the fifth-harmonic case. In addition, we found more significant differences in the amplitude envelope in the weak tails of the pulse ($\vert t \vert >$\SI{75}{\femto\second}), shown in Fig \ref{fig:data}a,c.

To understand the field sensitivity of our measurement, we calculated the field strength of the signal field based on its mode size diameter of \SI{134}{\micro\meter} and the pulse energy of \SI{1.25}{\nano\joule}, which is approximately \SI{100}{\mega\volt\per\meter}. Given the noise floor of the electric field measured \SI{1}{\pico\second} before the main feature of the pulse of \SI{60}{\decibel} we calculate a corresponding field noise floor of \SI{100}{\kilo\volt\per\meter}. 

To further benchmark our results, we Fourier transformed the measured fields and compared the spectral intensities against a measurement with a SWIR optical spectrum analyzer (OSA) of the identical signal beam (see Fig. \ref{fig:data} b.). 

We find significant agreement between the OSA and the fifth-harmonic based method and even find agreement in the short wavelength range from \SI{1.2}{\micro\meter} to \SI{1.7}{\micro\meter}, where we find a significant spectral plateau several orders of magnitude above the noise floor. The third-order method does not reproduce this spectral component on the short wavelength side, due to the smaller bandwidth of its transfer function. In the OSA measurement we find a drop in signal response on the long wavelength side, close to \SI{2.5}{\micro\meter} wavelength, due to the limited response of InGaAs detector in this wavelength range. However, both sampling methods show significant spectrum beyond the range of the OSA. The fifth-order method yields higher amplitude than the third-order method on the long wavelength shoulder, due to the increased bandwidth associated with the higher nonlinearity. In comparison, the spectra agree in their general shape also with the retrieved d-scan spectrum, but disagree in many of the more detailed spectral features which are only reproduced by the OSA. 

Overall, the field sampling approach yields an SNR of more than 6 orders of magnitude, limited by electrical readout noise, and surpasses the performance of the OSA measurement, while yielding a larger accessible bandwidth. It is also noteworthy that the d-scan seems to retrieve the overall pulse shape in the time domain, but many nuanced features of the pulse far away from its peak are lost. This can also be seen by the reconstructed spectral amplitude, which does not match qualitatively the results obtained with the OSA or our cross-correlations. It is unclear if the qualitative disagreements of the d-scan results are due to iterative retrieval noise or if there are other relevant noise limitations, highlighting the benefits of a direct electric field measurement technique. 

When comparing the measured spectral phase, we can also find excellent agreement of the d-scan phase with the ICC spectral phases, with only small deviations. Interestingly, a particular feature that is entirely missed by the d-scan retrieval is around \SI{1.5}{\micro\meter} in wavelength, where the spectral phase has a steep slope towards negative phase values. This spectral phase lies in a spectral region where the d-scan has not retrieved any spectral components beyond noise and therefore it is not expected that the spectral phase would be reliable.

\begin{figure}[h]
    \centering
    \includegraphics[width=16cm]{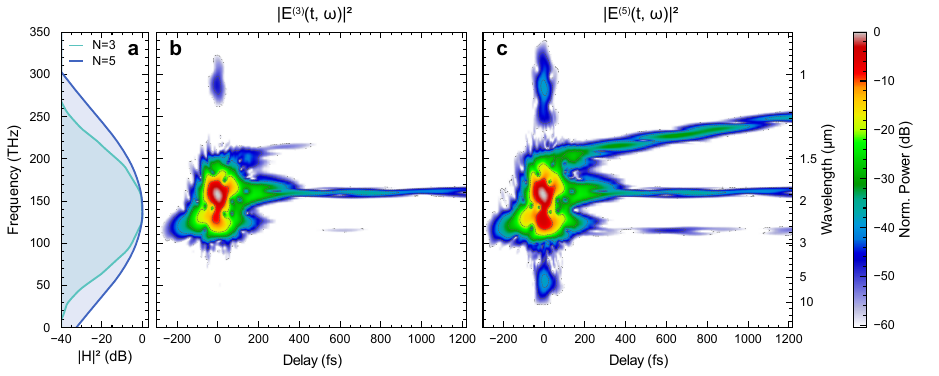}
    \caption{\textbf{Time-Frequency Analysis of Cross-Correlation Measurement Results: a,} Transfer function magnitude from Fig.~\ref{fig:transfer}. \textbf{b,} Spectrogram showing the time dependent frequency distribution of electric field measurement shown in \ref{fig:data} a. The spectrogram is calculated with an adaptive time window, ranging from ${\sim}\,\SI{100}{\femto\second}$ to ${\sim}\,\SI{400}{\femto\second}$ with a soft transition at ${\sim}\,\SI{250}{\femto\second}$. \textbf{c,} Spectrogram showing the time dependent frequency distribution of electric field measurement shown in \ref{fig:data} b. The spectrogram is calculated with an adaptive time window, ranging from ${\sim}\,\SI{100}{\femto\second}$ to ${\sim}\,\SI{400}{\femto\second}$ with a soft transition at ${\sim}\,\SI{250}{\femto\second}$. }
    \label{fig:spectrogram}
\end{figure}

To better visualize these spectro-temporal features in the measured electric field, we choose a time-frequency representation through the short-time Fourier-transform (STFT), that maps spectral components to their occurrence in time, see Fig. \ref{fig:spectrogram}. The STFT effectively slides a temporal window over the time domain of the measured pulse and takes a Fourier transform at every window position. We use an adaptive window size, that transitions from a \SI{100}{\femto\second} window to a \SI{400}{\femto\second} window at $t=$\SI{250}{\femto\second}. The adaptive window enables high temporal resolution capture of the \SI{100}{\femto\second} transient dynamics in the main pulse, as well as high spectral resolution capture of long-lived trailing features. The results of this analysis for the third- and fifth-harmonic approach are shown in Fig. \ref{fig:spectrogram}. In addition, we also plotted the spectral transfer function of the two methods as shown in Fig. \ref{fig:transfer}, to directly reference the spectrograms with the accessible bandwidth.

The spectrograms in Fig. \ref{fig:spectrogram} (b,c) of the third- and fifth-order methods reveal a rich spectro-temporal structure of the pulse that could not be easily understood from the decoupled time and frequency analysis plots. Both spectrograms show a central feature with significant trailing spectral components for more than \SI{1}{\pico\second}.
Around $t=0$ \SI{}{\femto\second} both spectrograms resemble the same shape and overall structure. Additionally, one can also observe indications of harmonic distortion occurring at \SI{300}{\tera\hertz} that is 4 to 5 orders of magnitude below the peak of the signal and is also spectrally well separated from the relevant components of the spectrogram. The transfer function shown in Fig. \ref{fig:spectrogram} a, also strongly indicates that this is purely distortion and can be neglected or further filtered out.

On the long wavelength (low frequency side) we also observe a spectral feature for the fifth-order method, which is not present in the third-order ICC spectrogram.
It is unclear if this is indeed a real spectral feature, which in principle would be supported by the transfer function bandwidth, or if it is the DC component of the distortion of the main pulse. Harmonic distortions can only be caused by strong fields, and the temporal position coincides with the main pulse, therefore strongly suggesting a harmonic distortion as a reasonable explanation.

The revealing features of the spectrograms are the trailing dynamics starting at $t> $\SI{200}{\femto\second} and are detected by both approaches. We can identify three distinct spectral bands, the first band at \SI{115}{\tera\hertz}, the second at \SI{160}{\tera\hertz} and a broadband band ranging from \SI{200}{\tera\hertz} to \SI{250}{\tera\hertz} with a strong spectral chirp.
Only the fifth-order method is able to fully capture these three distinct trailing spectral bands, as the third-order method does have sufficient bandwidth, causing significant attenuation of the low and the high frequency band to the point that they become indistinguishable from noise. The broadband high frequency band is a residual component from our OPA based laser source, that was in the past never observed with conventional pulse characterization methods, such as d-scan. The strong spectral chirp is caused by the dispersion compensation of the system, that only compresses spectral components between \SI{100}{\tera\hertz} and \SI{200}{\tera\hertz}. Details on the compression and chirped mirror design are found in reference \cite{ritzkowskyHighrepetitionRateCEPstable2026}.

\begin{figure}[h]
    \centering
    \includegraphics[width=16cm]{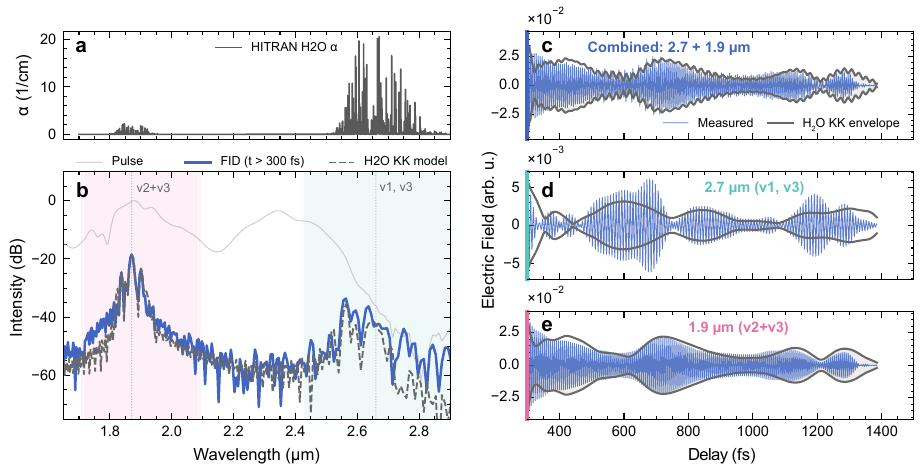} 
    \caption{\textbf{Analysis of the Free-Induction Decay of Water: a,} Absorption cross-section as a function of wavelength provided by the HITRAN database. \textbf{b} spectral intensity of the time-windowed (t > \SI{300}{\femto\second}) Fourier transformed free-induction decay (FID) from the measured data with $n=5$ (blue) and a simulated FID based on the HITRAN data (grey dashed). The prompt pulse response (time window t < \SI{300}{\femto\second}) is shown in light grey for comparison. \textbf{c,} Time domain electric field of the frequency filtered FIDs of the \SI{1.9}{\micro\meter} and the \SI{2.7}{\micro\meter} water absorption band comparing the measured data against the simulated response shown as an envelope. \textbf{d,} Frequency filtered time-domain FID response of the symmetric ($\nu_1$) and asymmetric O-H bond stretch ($\nu_3$) of the water molecule. \textbf{e,} Frequency filtered time-domain FID response of the \SI{1.9}{\micro\meter} combination band of the H-O-H bending mode ($\nu_2$) and the asymmetric O-H bond stretch ($\nu_3$).}
    \label{fig:fid_analysis}
\end{figure}

The other two spectral bands result from free-induction decay (FID) of ambient water vapor absorption lines\cite{srivastavaNearpetahertzFieldoscopyLiquid2024}. They are caused by excitation of ro-vibrational modes of the H$_2$O molecule, with three distinct vibrational modes contributing. At a low frequency around \SI{2.7}{\micro\meter}, we detect the symmetric O-H stretch mode $\nu_1$ and  the asymmetric O-H stretch mode $\nu_3$. At \SI{1.9}{\micro\meter} we detect the combination band of the $\nu_3$ and the H-O-H bend mode $\nu_2$. To demonstrate that the measured field is quantitatively accurate, we compare it against a calculation of the expected FID from ambient water vapor interacting with our laser pulse. With H$_2$O absorption data provided by the HITRAN database, we calculate the phase response of the FID by applying the Kramers-Kronig relation to the reference absorption spectrum \cite{gordonHITRAN2020MolecularSpectroscopic2022}, see Fig. \ref{fig:fid_analysis} a. For the FID model we use the measured relative humidity of \SI{14.9}{\percent}, resulting in a volume mixing ratio of \SI{0.481}{\percent}, and a measured temperature of \SI{25.5}{\celsius}, as well as a measured \SI{6.1}{\meter} of free propagation. We take a windowed version of the pulse ($t<300$ \SI{}{\femto\second}) to get the initial pulses the electric field without the absorption features, see Fig. \ref{fig:fid_analysis} b. The modeled FID signal can be calculated by applying the impulse response of the water absorption lines to the initial pulse component. Spectrally windowing both the calculated and the measured FID, between \SI{300}{\femto\second} and \SI{1400}{\femto\second}, allows a fair spectral comparison. 

The Fourier-transformed FID tail is shown in Fig. \ref{fig:fid_analysis} b. Around the \SI{1.9}{\micro\meter} band we find a slightly broader structure, which is due to a small spectral bleed of the prompt pulse into the tails, which is not present in the calculated data. In the \SI{2.7}{\micro\meter} band we find stronger deviations from the calculated data, stemming from less SNR compared to the other band. We emphasize that our FID field calculations use no fitting parameters. The close agreement between the measured and calculated FID response highlights the quantitative accuracy of our direct measurement technique for self-referenced field-resolved spectroscopic applications.

For better visualization, we spectrally isolate the two individual bands and plot their temporal FID electric fields in Fig. \ref{fig:fid_analysis}c,d,e. We find that also in the time domain we have significant agreement between the measured and the calculated FID. The comparison between the two bands shows that the \SI{2.7}{\micro\meter} band is almost one order of magnitude smaller in electric field strength than the other band and shows some deviations from the calculated response. However, the \SI{1.9}{\micro\meter} band has excellent agreement with the calculated response, highlighting on a quantitative level the performance of asymmetric ICC for self-referenced field-resolved detection. It is noteworthy that most of the absorption occurred on the beam path before the ICC interferometer and that this emphasizes the absolute measurement of the electric field in contrast to conventional homodyne techniques, which would only measure spectral differences between two arms of an interferometer. To further explore the \SI{2.7}{\micro\meter} absorption band we deliberately broadened the signal pulse through self phase modulation in TiO$_2$. We found a significant increase in SNR of this band and also show that the spectral range of the fifth-harmonic method allows to fully detect the broadened pulse covering \SI{1.2}{\micro\meter} up to \SI{3.5}{\micro\meter}, see Extended Data Fig. \ref{fig:SuppFID} for the results.

\section{Conclusion}

We have demonstrated all-optical nonlinear interferometric cross-correlation as a quantitative tool for field-resolved spectroscopy approaching the performance of state of the art EOS\cite{srivastavaNearpetahertzFieldoscopyLiquid2024}. The complexity of our field sampling approach is significantly reduced in comparison to other ionization based interferometric cross-correlation schemes or EOS and only requires a thin-film nonlinear medium and a balanced Mach-Zehnder interferometer. Driving fifth-harmonic generation in a 150-nm ITO film, we achieved a measurement bandwidth of 190~THz (80--270~THz) at sub-500~GHz spectral resolution and a temporal and spectral intensity dynamic range exceeding \SI{60}{\decibel} down to field strength of \SI{100}{\kilo\volt\per\meter}. These capabilities allowed us to resolve the free-induction decay of two ro-vibrational bands of ambient water with parameter-free agreement against HITRAN. Because the absorption accumulates along the beam path before the interferometer, we emphasize that the measurement is sensitive to the complex field in amplitude and phase, rather than to a differential amplitude between two arms, as would be common in typical homodyne detection schemes.

By leveraging concepts and figures of merit long-established within the microwave community, in particular IMD and SFDR, we identify an operating regime in which faint spectral features are unambiguously differentiated from the distortion floor without the need for otherwise common iterative retrieval algorithms\cite{choReconstructionAlgorithmTunneling2021,wieseUniversalWaveformresolvingDual2024,leungUniversalInlineWaveform2026}. This is a critical distinction from previous demonstrations of asymmetric IAC, that relied on amplitude-weighted reconstruction procedures that are structurally biased against the weak spectral components of greatest spectroscopic interest \cite{gliserinCompleteCharacterizationUltrafast2022,choReconstructionAlgorithmTunneling2021,wieseUniversalWaveformresolvingDual2024}. We emphasize that these concepts and metrics are agnostic to the readout modality: they apply equally to the all-optical scheme we demonstrate in this work as well as to current-based variants based on tunnel ionization~\cite{parkDirectSamplingLight2018,biontaOnchipSamplingOptical2021,yeungLightwaveelectronicHarmonicFrequency2024} or multiphoton absorption~\cite{liuSingleshotMeasurementFewcycle2022,liuAllopticalSamplingFewcycle2021}, and in principle to any nonlinear element whose response can be characterized in terms of its harmonic content.  As such, we argue they should be leveraged across the community to provide standard quantitative performance benchmarks across techniques that can be used to guide the future development  of petahertz field-resolved spectroscopy
\cite{herbstRecentAdvancesPetahertz2022,heidePetahertzElectronics2024}.  By supplying common figures of merit across techniques and laboratories, the mixer framework provides the quantitative basis on which field-resolved spectroscopy using nonlinear cross-correlations can mature from a collection of pulse characterization demonstrations into a standardized tool of ultrafast science \cite{pupezaFieldresolvedInfraredSpectroscopy2020,huberStandardizedElectricFieldResolvedMolecular2024,hoferLinearFieldresolvedSpectroscopy2025}.

Given the performance and simplicity of the all-optical nonlinear ICC approach we present here, we anticipate  rapid adoption and broad impact beyond the ultrafast optics research community. 
In particular, we see direct applications in shortwave and mid-infrared stand-off trace-gas spectroscopy\cite{ycasMidinfraredDualcombSpectroscopy2019,riekerFrequencycombbasedRemoteSensing2014a} or, more importantly, in translating the all optical-sampling scheme to shorter wavelength to enable an elegant method for field resolved detection of correlated emission in quantum materials in the near-infrared \cite{yeungFieldresolvedObservationExciton2026}.

\section{Methods}\label{Methods:Setup}

The laser source used for the experiments is a carrier-envelope-phase-stable shortwave-infrared source generating few-cycle pulses centered at 2060 nm, described extensively in an earlier publication \cite{ritzkowskyHighrepetitionRateCEPstable2026}. The system is pumped by a 1030 nm Yb:KGW amplifier delivering 220 fs pulses at 400 kHz. A small fraction of the pump is used to generate a near-infrared white-light continuum in YAG, which is mixed with the 1030 nm pump in a type-I BBO crystal to produce a CEP-stable difference-frequency seed. This seed is pre-compressed with chirped mirrors and amplified through two non-collinear BiBO optical parametric amplifier stages operated near degeneracy. The amplified SWIR pulses are then compressed using custom chirped mirrors and ZnSe wedges, with residual near-infrared and visible light removed by a thin silicon filter. The final output delivers 13.6 fs, two-cycle pulses with 2.25 µJ energy at 400 kHz, suitable for field-resolved spectroscopy.

The output of the SWIR laser source was directed into an interferometric setup consisting of a signal arm and a gate arm~\ref{fig:setup}. Each arm included ZnSe wedges for independent dispersion tuning. In the gate arm, the pulse energy was adjusted using a motorized broadband half-wave plate followed by a wire-grid polarizer. The measurement was performed stroboscopically by scanning the relative delay with a voice-coil delay stage, oscillating \SI{2}{\hertz} and with a delay range of over \SI{2.5}{\pico\second}, placed in the gate arm. The position of the voice-coil was tracked optically to account for instabilities. The signal arm goes through a de-magnifying intermediate focus with concave metal mirrors of \SI{15}{\centi\meter} and \SI{25}{\centi\meter} focal length, and a variable aperture with a diameter of \SI{>0.5}{\milli\meter} to reduce the transmitted energy and the collimated mode size. The output average power of the signal is on the order of \SI{500}{\micro\watt} and with a mode size of \SI{0.5}{\milli\meter}. Both signal and gate are collinearly guided and focused with a \SI{25}{\milli\meter} focal length metallic off-axis parabola at the commercial sputter-coated \SI{150}{\nano\meter} thick ITO target on a \SI{3}{\milli\meter} thick BK7 substrate. The reduced collimated mode size of the signal, will yield a relative attenuation of \SI{0.02}{} of the signals peak power in the focus, due to the larger focal mode diameter of \SI{134}{\micro\meter} in contrast to the original diameter of \SI{19}{\micro\meter}. This gives a relative attenuation to the gate with \SI{8}{\milli\watt} average power in the n~=~3 case of 0.00125 in the focus and for the n~=~5 case, where gate average power is \SI{80}{\milli\watt}, where the relative attenuation is 0.000125. The field attenuation factor $s$ is therefore $s_\text{n=3}=0.035$ and $s_\text{n=3}=0.011$. After the nonlinear signal generation we used a combination of a KG filter and a shortpass filter to block the residual fundamental and selected the harmonics of interest (predominantly the third and the fifth), which were then measured with a commercial silicon photodiode.

We acquired 800 individual ICC signal interferograms with a 12-bit ADC, with simultaneous acquisition of the position tracking signal from the voice-coil delay line. This approach ensured a fully referenced delay position across the interferogram signal. Using the delay stage position reference, the individually acquired interferograms were coherently averaged to reduce noise. The averaged traces were spectrally filtered to remove residual noise outside the relevant measurement bandwidth and to reduce harmonic distortions at DC and the second harmonic of the signal. 

\bibliography{references}

\begin{backmatter}
\bmsection{Funding}
This work was supported by the U.S. Department of Energy, Office of Science, Office of Basic Energy Sciences, under award no. DE-SC0024173. F. R. acknowledges support from the Alexander von Humboldt Foundation and the European Research Council under the ERC SoftMeter no. 101076500.  G. L. D. acknowledges support from the Progetto Rocca Postdoctoral Fellowship. B. M. M. acknowledges support from the NDSEG Fellowship. M. Y. acknowledges support from the National Science Foundation MPS-Ascend Postdoctoral Research Fellowship under grant no. 2402151. 

\bmsection{Acknowledgment}
The authors thank Zhenyang Xiao and Karl K. Berggren for providing helpful feedback on this manuscript. FR acknowledges the scientific support of Andrea Trabattoni for the writing process of this manuscript and fruitful scientific discussions.

\bmsection{Disclosures}
The authors declare no conflicts of interest.

\bmsection{Data Availability Statement}
The data and analysis code is available at the following data repository: \url{https://github.com/ritzkowsky42/All-Optical-Field-Resolved-Spectroscopy-With-Interferometric-Nonlinear-Cross-Correlations}.

\bmsection{Supplemental document}
See Supplement 1 for supporting content.

\bmsection{Extended Data Figures}

\begin{figure}[h]
    \centering
    \includegraphics[width=16cm]{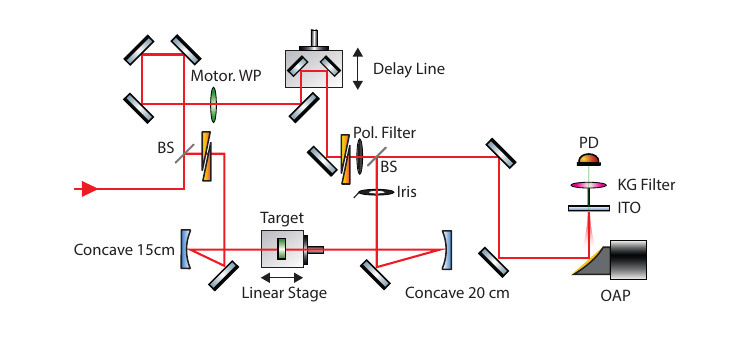}
    \setcounter{figure}{0}
    \renewcommand{\figurename}{Extended Data Fig.}
    \renewcommand{\thefigure}{\arabic{figure}}
    \caption{\textbf{Interferometric field-resolved spectroscopy setup.} The output of the CEP-stable SWIR source is split into a signal arm and a gate arm. Both arms include ZnSe wedges for independent dispersion tuning. In the gate arm, a motorized broadband half-wave plate (WP) and wire-grid polarizer (Pol. filter) control the pulse energy, while an oscillating voice-coil delay stage enables stroboscopic scanning of the relative delay. In the signal arm, an iris controls the pulse energy before the beam is focused into the interaction region, where it can interact with a sample or target and undergo nonlinear optical processes such as self-phase modulation. The signal and gate beams are then recombined at a second beamsplitter (BS). An 25-mm-focal off-axis metallic parabolic mirror (OAP) focuses the beams onto the ITO, and a filter selects the third and fifth harmonic, collected by a Silicon photo-diode (PD).}
    \label{fig:setup}
\end{figure}

\begin{figure}[h]
    \centering
    \includegraphics[width=8cm]{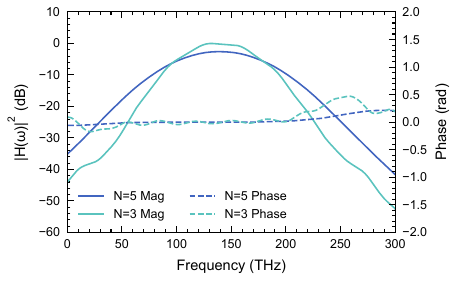}
    \renewcommand{\figurename}{Extended Data Fig.}
    \renewcommand{\thefigure}{\arabic{figure}}
    \caption{\textbf{Sampling Transfer Function:} Gate magnitude and phase response as a function of frequency and nonlinearity. The effective gate transfer function is calculated based on the nonlinearity of the system and the measured electric field of the gate characterized via d-scan.}
    \label{fig:transfer}
\end{figure}

\begin{figure}[h]
    \centering
    \renewcommand{\figurename}{Extended Data Fig.}
    \renewcommand{\thefigure}{\arabic{figure}}
    \includegraphics[width=16cm]{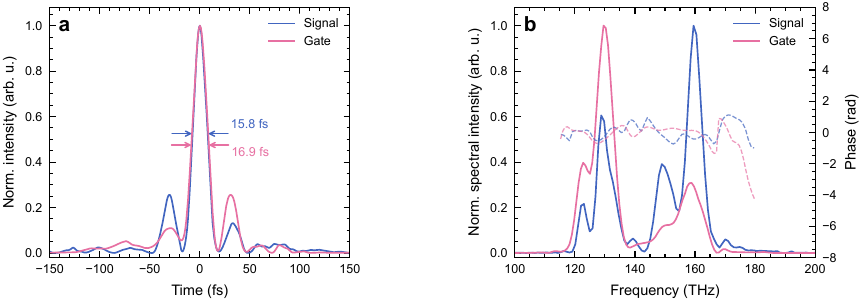}
    \caption{\textbf{Dispersion-Scan Characterization: a,} Dispersion scan characterized temporal envelopes of the signal and the gate arm at an incident pulse energy of around \SI{20}{\nano\joule}. The dispersion scan was realized through third-harmonic generation on an ITO film and the nonlinear signal was detected with a Silicon spectrometer. The retrieval algorithm is based on the l-bfgs-b nonlinear optimizer. \textbf{b,} Spectral amplitude (solid lines) and spectral phase (dashed lines) of the retrieved pulses shown in a.}
    \label{fig:dscan}
\end{figure}

\begin{figure}[h]
    \centering
    \renewcommand{\figurename}{Extended Data Fig.}
    \renewcommand{\thefigure}{\arabic{figure}}
    \includegraphics[width=8cm]{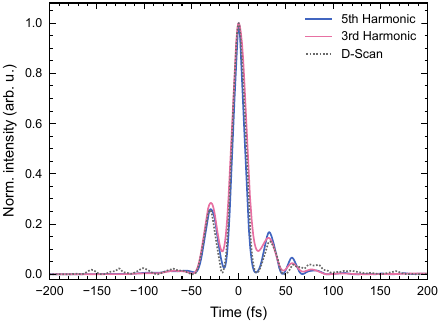}
    \caption{\textbf{Time-Domain Comparison of the Characterized Fields:} Comparison of the ICC measured intensity envelopes with the measured dispersion scan result. The fifth-harmonic ICC measures a \SI{13.5}{\femto\second} duration pulse, whereas the third-harmonic measures \SI{17.4}{\femto\second} in duration. The dispersion scan method retrieved a duration of \SI{15.8}{\femto\second}.}
    \label{fig:comparison}
\end{figure}

\begin{figure}[h]
    \centering
    \renewcommand{\figurename}{Extended Data Fig.}
    \renewcommand{\thefigure}{\arabic{figure}}
    \includegraphics[width=10cm]{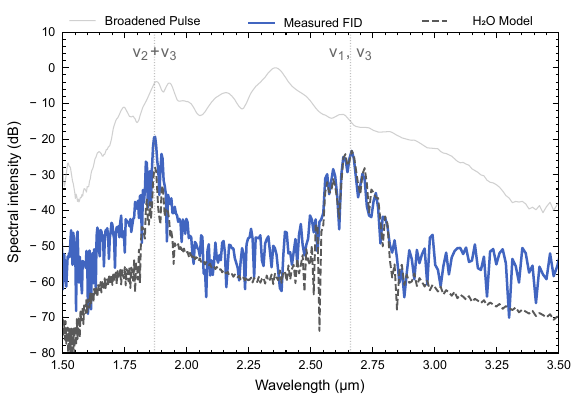}
    \caption{\textbf{Supplementary Analysis of the Free-Induction Decay in spectrally broadened signal:} Spectral intensity of the time-windowed (t > 300 fs) Fourier-transformed FID from the spectrally broadened signal pulse (blue) and simulated FID based on the HITRAN data(grey dashed). The prompt signal spectrum is shown in grey.
    \label{fig:SuppFID}}
\end{figure}

~\newline

\end{backmatter}

\clearpage


\section*{Supplementary material}

## 1. Data Analysis Procedure

To acquire the datasets shown in the main text of this manuscript, we followed the data analysis strategy shown in the following paragraphs.

The measured photodiode signals were acquired with an oscilloscope on a 2 Hz trigger, that was provided by the voice coil linear stage oscillating with 2 Hz, shown in Fig. S1a. Similarly, we acquired a reference signal from an auxiliary interferometer that uses a 975 nm single-mode laser diode to provide a position reference for the moving voice coil stage Fig. S1a and b. Then we proceeded to cut the single trigger signal into the up and down ramp motion of the voice coil stage, as shown in Fig. S1c.

The next step in the analysis procedure is to coherently average all acquired interferograms. For this we separately average the up and the down ramp and suppress technical measured in the laboratory timescale of the interferograms on the kHz scale. This process is shown in Fig. S2, where we apply a passband filter from 0.4 kHz to 12 kHz. In this spectrum on the kHz-scale one can already identify the measured 2.1 ţm spectrum between 2 kHz and 4 kHz. The filtered signal is then used to coherently average all the individual interferogram. Between individual interferograms we found a few femtosecond jitter that was corrected by calculating the cross-correlation between the individual interferogram and the calculated average. The calculated average represents the average of all previous interferograms and gets updated continuously while analyzing all interferograms.

m

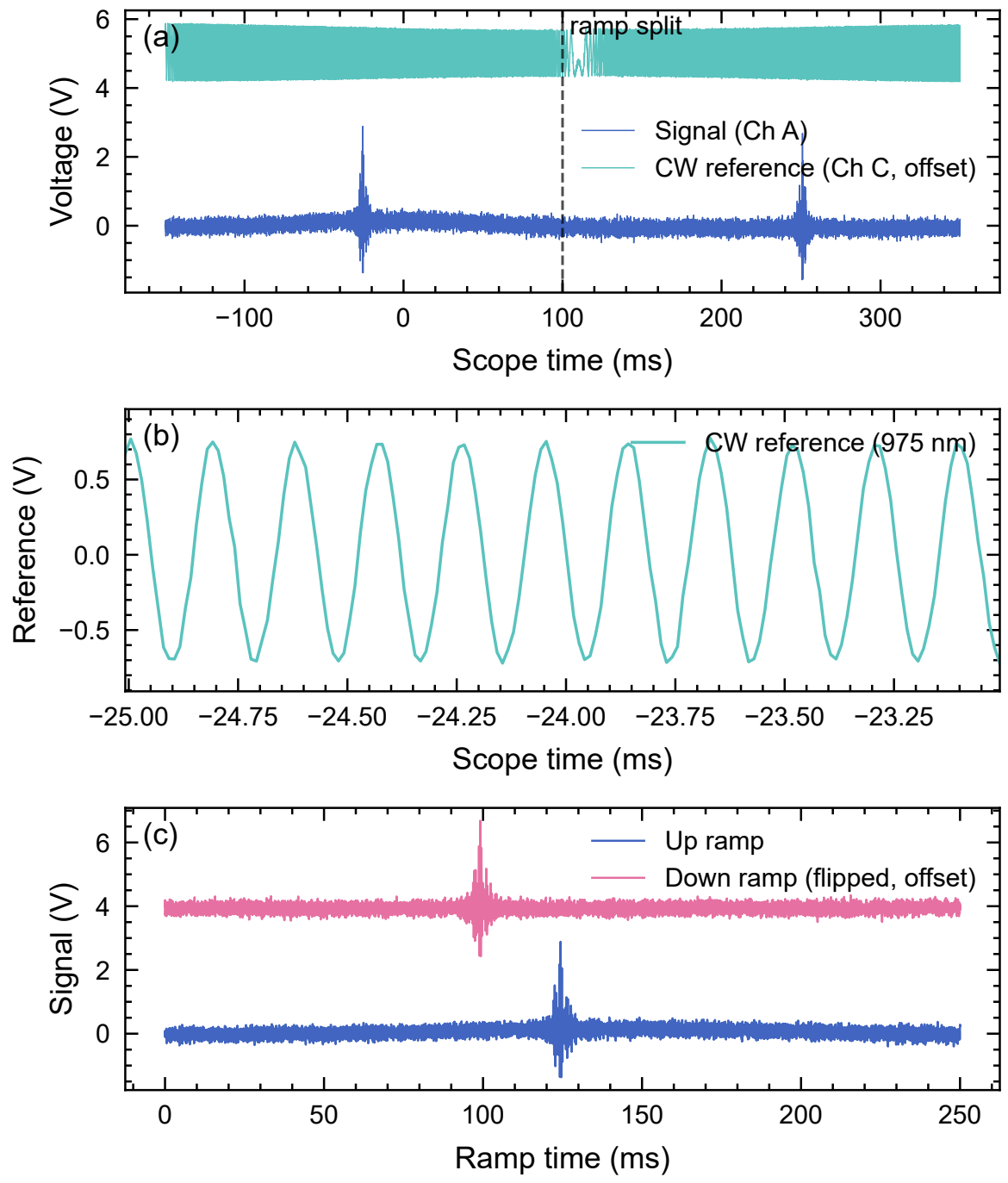


**Fig. S1. Raw Signal Acquisition and Interferometric Delay Reference: a**, We show the raw signals acquired on a 2 Hz basis with the sampling signal in blue and the delay reference in green. The raw signals are split into the up and the down ramp of the delay position as indicated by the black dashed line. **b**, Zoom in of the delay reference signal produced by the auxiliary interferometer and a 975 nm single-mode laser diode. **c**, the individual raw interferogram traces corresponding to the up and the down ramp interferogram produced by the oscillating linear stage.

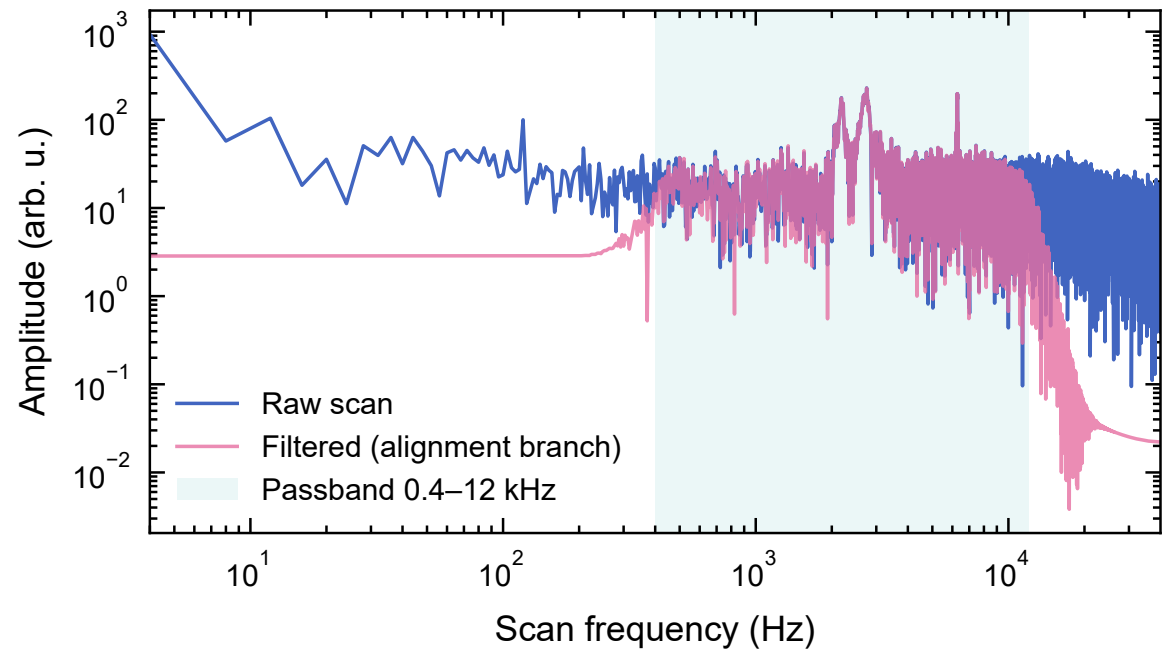


**Fig. S2. Frequency Representation of the Raw Measurement Signal:** Fourier transformed raw interferogram on the laboratory frequency scale and the filtered signal used for alignment of the individual traces.

The resulting averaged signal is shown in Fig. S3 a and b. In addition to the filtering of technical noise on the laboratory frequency scale, we also a apply a passband from 50 THz to 250 THz on the optical frequency scale to remove the 2nd order distortion effects at DC and at the second harmonic of the measured signal. The effect of this filter is shown in Fig. S3a. The final averaged waveform signal is shown in Fig. S3b in units of volts. The volts amplitude corresponds to the signal level measured on the photodiode.

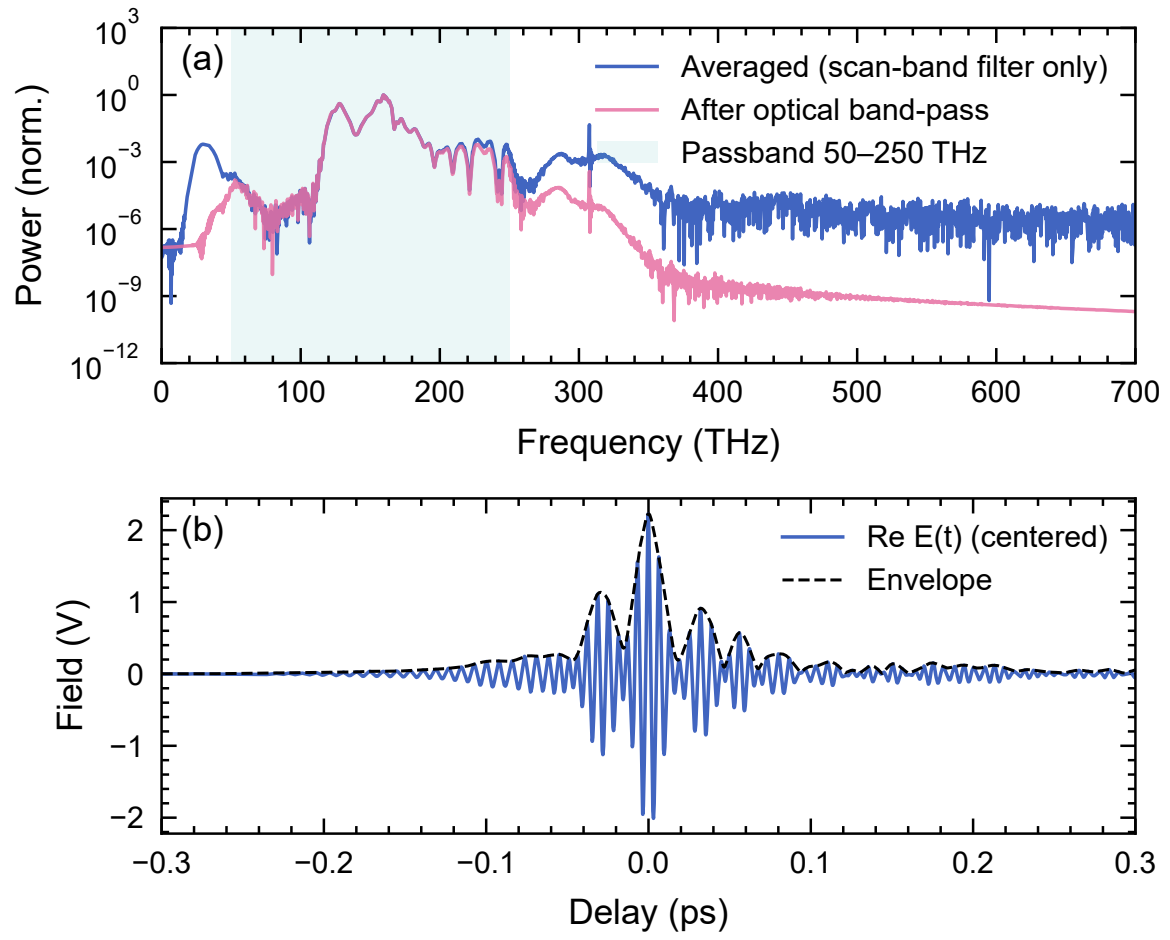


**Fig. S3. Averaged Measurement Signal in Frequency and Time: a** Frequency representation of the averaged measurement signal and the passband filtered signal to suppress the 2nd order distortion effects. **b**, time domain signal of the averaged measured waveform shown in units of Volts of the photodiode signal.

## 2. Analytical Derivation of Nonlinear Distortion Artifacts

Here we will analyze the theory of nonlinear cross-correlations with arbitrary order perturbative harmonics such as second-harmonic generation (SHG), third-harmonic generation (THG), or high-harmonic generation (HHG). This derivation follows from Gliserin et al. [1].

In a nonlinear cross-corelation, there is a strong gate field $E_G(t)$, which drives the nonlinear response and a weak signal field $E_S(t-\tau)$, which is the desired field to measure. The signal pulse is attenuated with a pre-factor $s < 1$ and collinearly recombined with the gate pulse. The pulses are guided onto a medium with a nonlinear susceptibility $\chi^{(n)}$ of order $n$, which generates the nth-harmonic of the gate field. An optical filter isolates this harmonic, which is then detected by a square-law photodetector. The detector records the resulting nonlinear cross-correlation waveform $I_{\mathrm{CC}}^{(n)}(\tau)$ as a function of time delay $\tau$,

$$I_{\mathrm{ICC}}(\tau) \propto \int \left| \left(E_G(t) + sE_S(t-\tau)\right)^n \right|^2 dt. \tag{S1}$$

Using the binomial identity,

$$(x+y)^n = \sum_{m=0}^{n} \binom{n}{m} x^m y^{n-m}, \tag{S2}$$

we can rewrite the cross-correlation function as:

$$I_{\mathrm{ICC}}(\tau) \propto \int \left| \sum_{m=0}^{n} \binom{n}{m} E_G^{n-m}(t)\, s^m\, E_S^m(t-\tau) \right|^2 dt. \tag{S3}$$

Expanding the first few terms shows the general structure of the equation[1]:

$$\begin{aligned} I_{\mathrm{ICC}}(\tau) \propto \int \Bigg| & \binom{n}{0} E_G^n(t) + \binom{n}{1} s\, E_G^{n-1}(t)\, E_S(t-\tau) \\ & + \binom{n}{2} s^2\, E_G^{n-2}(t)\, E_S^2(t-\tau) + \binom{n}{3} s^3\, E_G^{n-3}(t)\, E_S^3(t-\tau) + \ldots \Bigg|^2 dt. \end{aligned} \tag{S4}$$

We define the term $A_m(t,\tau) = \binom{n}{m} s^m\, E_G^{n-m}(t)\, E_S^m(t-\tau)$ to make the integrated modulus squared more compact,

$$\left| \sum_{m=0}^{n} A_m(t,\tau) \right|^2 = \sum_{m=0}^{n} \sum_{\ell=0}^{n} A_m(t,\tau) A_\ell^*(t,\tau). \tag{S5}$$

The pre-factor $s < 1$ sets the relative weight of the individual components of this sum, would be enumerated with $s^{m+\ell}$. Before evaluating the modulus squared, we note that each factor of $E_S(t-\tau)$ carries a carrier phase $e^{-i\omega_0\tau}$ from the delay. Consequently, the $m$th-order term in the expansion oscillates as $e^{-i\omega_0\tau\cdot m}$ in $\tau$. After squaring, cross-terms between orders $m$ and $\ell$ oscillate at $(m-\ell)\omega_0$. This frequency separation is crucial and allows us to isolate the signal. Furthermore, it determines which distortion terms are separable by Fourier filtering.

In the following derivation, we keep track of the power of $s$ and the oscillation frequency in $\tau$ (determined by the polynomial order of $E_S$), retaining all terms up to $\mathcal{O}(s^3)$.

There is one term at $s^0$ (DC in $\tau$, i.e. $m - \ell = 0$):

$$\left|E_G^n(t)\right|^2. \tag{S6}$$

There is one configuration at $\mathcal{O}(s^1)$, which constitutes our signal ($m - \ell = \pm 1$, oscillating at $\omega_0$ in $\tau$):

$$\binom{n}{1} s\, [E_G^*(t)]^n \cdot E_G^{n-1}(t) \cdot E_S(t-\tau) + \text{c.c.} = \binom{n}{1} s\, E_G^*(t) \cdot |E_G(t)|^{2(n-1)}(t) \cdot E_S(t-\tau) + \text{c.c.} \tag{S7}$$

The second-order distortion receives contributions from two products: the $\mathcal{O}(s^2)$ term with the zeroth-order conjugate, and the product of the first-order term with its own conjugate. These yield a second-harmonic distortion and a DC distortion, respectively:

$$\underbrace{\binom{n}{2} s^2\, [E_G^*(t)]^n\, E_G^{n-2}(t)\, E_S^2(t-\tau) + \text{c.c.}}_{\text{2nd-order harmonic distortion, at } 2\omega_0\tau} + \underbrace{\binom{n}{1}^2 s^2 \left|E_G^{n-1}(t)\right|^2 |E_S(t-\tau)|^2}_{\text{DC distortion (real, no c.c. needed)}}. \tag{S8}$$

The first term is complex and oscillates at $2\omega_0$ in $\tau$; the second is a product of modulus squares and contributes only at DC in $\tau$. Both terms are **separable** from the signal: the $2\omega_0\tau$ component is removed by Fourier filtering, and the DC contribution is removed by background subtraction. This explains the harmonic distortion observed in the spectrograms at $\sim 300\,\mathrm{THz} \approx 2\omega_0$ (Fig. 4), which lies 45 orders of magnitude below the peak signal and is spectrally well separated.

[1] Binomial coefficients: for $n = 5$: $1, 5, 10, 10, 5, 1$; for $n = 3$: $1, 3, 3, 1$.

The third-order distortion involves additional terms. We obtain the product of $\mathcal{O}(s^3)$ with the zeroth-order conjugate ($m - \ell = 3$, oscillating at $3\omega_0\tau$), and the product of the second- and first-order terms ($m - \ell = 1$, oscillating at $\omega_0\tau$):

$$\begin{aligned}&\binom{n}{3} s^3 \, [E_G^*(t)]^n \, E_G^{n-3}(t) \, E_S^3(t-\tau) \\ &+ \binom{n}{2} s^3 \binom{n}{1} \left(E_G^{n-2}(t) \, E_S^2(t-\tau)\right) \cdot \left(E_G^{n-1}(t) \, E_S(t-\tau)\right)^* + \text{c.c.,}\end{aligned} \tag{S9}$$

which can be rewritten as:

$$\begin{aligned}&\underbrace{\binom{n}{3} s^3 \, [E_G^*(t)]^n \, E_G^{n-3}(t) \, E_S^3(t-\tau)}_{\text{3rd-order harmonic distortion, at } 3\omega_0\tau} \\ &+ \underbrace{\binom{n}{2}\binom{n}{1} s^3 \, E_G^*(t) \, \left|E_G^{n-2}(t)\right|^2 E_S(t-\tau) \, |E_S(t-\tau)|^2}_{\text{3rd-order intermodulation distortion, at } \omega_0\tau} + \text{c.c.}\end{aligned} \tag{S10}$$

The first term oscillates at $3\omega_0\tau$ and is separable by Fourier filtering. The second term, the intermodulation distortion, oscillates at $\omega_0\tau$, identical to the signal frequency. **It is inseparable from the signal by any filtering operation.** This is the critical distortion term that limits measurement fidelity. However, this distortion is present in many conventional electrical and RF systems and can be ignored under the right circumstances, which will be elaborated in the following. The intermodulation term is proportional to $E_S|E_S|^2$, meaning it contributes an effective signal that scales as the cube of the signal field, distorting the retrieved waveform in amplitude and phase. It also scales with the nonlinearity order through the binomial coefficients, which are generally larger for higher $n$; for $n = 5$ the intermodulation coefficient is $\binom{5}{2}\binom{5}{1} = 50$.

We can now deduce the relative scaling to the measured signal. The signal amplitude is $\binom{n}{1}s = ns$ and the intermodulation amplitude is $\binom{n}{2}\binom{n}{1}s^3 = 50s^3$ for $n = 5$. Taking the ratio:

$$\frac{\text{Intermodulation amplitude}}{\text{Signal amplitude}} = \frac{\binom{n}{2}\binom{n}{1} s^3}{\binom{n}{1} s} = \binom{n}{2} s^2 = \frac{n(n-1)}{2} s^2. \tag{S11}$$

For $n = 5$ this gives $10s^2$, which in intensity corresponds to $\binom{n}{2}^2 s^4 = 100s^4$.

Similarly, for the 2nd-order (harmonic) distortion:

$$\frac{\text{2nd-order amplitude}}{\text{Signal amplitude}} = \frac{\binom{n}{2} s^2}{\binom{n}{1} s} = \frac{n-1}{2} s = 2s \quad (n = 5). \tag{S12}$$

In our experiment the amplitude ratio is $s \approx 10^{-2}$, so the intermodulation product is:

$$100 \cdot s^4 = 100 \cdot 10^{-8} = 10^{-6} \quad (-60\,\text{dB in intensity}) \tag{S13}$$

below the signal, well beneath the detected noise floor. The 2nd-order harmonic distortion is larger in amplitude ($2s = 2 \times 10^{-2}$, i.e. $-34\,$dB in intensity), but oscillates at $2\omega_0\tau$ and is completely removed by the spectral filtering applied to the interferogram.

We define the signal level at which intermodulation distortion causes a 1 dB departure (approximately 12% error in amplitude) by setting $\binom{n}{2}s^2 \approx 0.12$:

$$s_{1\,\text{dB}} = \sqrt{\frac{0.12}{\binom{n}{2}}} = \sqrt{\frac{0.12}{10}} \approx 0.11 \quad (n = 5). \tag{S14}$$

For $n = 3$: $s_{1\,\text{dB}} = \sqrt{0.12/3} \approx 0.20$.

This means that for the fifth-harmonic method, the measurement remains faithful to within 1 dB as long as $s < 0.11$, i.e. the signal amplitude is less than approximately 11% of the gate amplitude. For our experimental conditions ($s \sim 10^{-2}$) we are more than an order of magnitude below this threshold.

This also implies that the higher-order channel ($n = 5$) enters the nonlinear regime before the lower-order channel ($n = 3$), since $s^{(5)}_{1\,\text{dB}} < s^{(3)}_{1\,\text{dB}}$. The agreement between the $n = 3$ and $n = 5$ retrieved fields (Fig. 3) therefore serves as an internal consistency check, confirming that both channels operate within their respective linear regimes.

## 3. Effects of Filter and Photodiode Spectral Response on Measured Signal

Here we analytically show the effects of the spectral response of components, downstream from the nonlinear medium, that measure $I_{\text{ICC}}^{(n)}$. These include the response of the n-th harmonic filter $H_f(\omega)$, the photodetector responsivity $R_{pd}(\omega)$, and any other optics $H_o(\omega)$. These can be combined into a single variable $H_m(\omega)$, which has a temporal response $h_m(t)$.

$$H_m(\omega) = H_f(\omega) R_{pd}(\omega) \prod_k H_{o,k}(\omega). \tag{S15}$$

Including $h_m(t)$ in S1,

$$I_{\text{ICC}}^{(n)}(\tau) \propto \int \left| \left(E_G(t) + sE_S(t-\tau)\right)^n * h_m(t) \right|^2 dt. \tag{S16}$$

By Parseval's theorem and the convolution theorem,

$$I_{\text{ICC}}^{(n)}(\tau) \propto \int \left| \mathscr{F}\left\{ \left[E_G(t) + sE_S(t-\tau)\right]^n \right\} \right|^2 \cdot \left| H_m(\omega) \right|^2 d\omega. \tag{S17}$$

Critically, this shows that due to the nonlinearity of the cross-correlation, the spectral phase of $H_m(\omega)$ has no effect on the fidelity of the output signal; the spectral phase of the signal of interest is retained in the nonlinear product. The output signal is only affected by the spectral magnitude $|H_m(\omega)|$. Any mismatch between the spectrum of $|H_m(\omega)|$ and the spectrum of the modulated harmonic introduces error.

In our experiment, this error would be dominated by the nonuniform responsivity of the photodiode reading out $I_{CC}^{(n)}$. The harmonic selection filter and all other optics preceding detection have a roughly flat spectral response over the relevant bandwidth. A Thorlabs DET36A photodetector was used to record $I_{CC}$. Its responsivity curve is shown below.

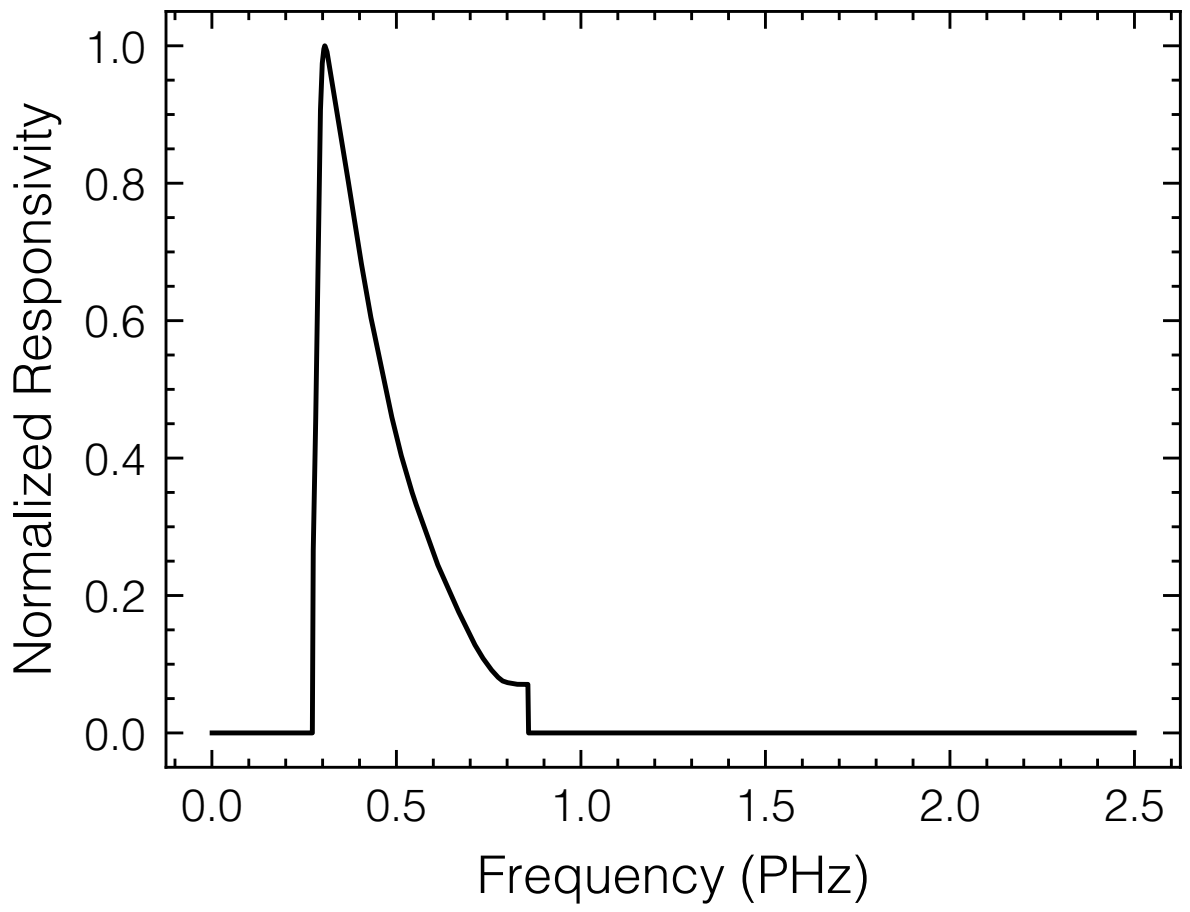


**Fig. S4. Thorlabs DET36A responsivity curve**

As a conservative estimate, the responsivity has been truncated to zero outside the wavelength limits of the provided manufacturer specifications. To investigate the error caused by the photodiode spectral response, we numerically simulated the experiment with the d-scan retrieved gate and signal fields. We compare the RMS errors of $I_{CC}^{(n)}$ assuming an ideal flat photodiode and $I_{CC}^{(n)}$ filtered by the photodiode spectral response. For both n=3 and n=5, we find that the added RMS error from the photodiode response is 0.5% in amplitude, which indicates that the photodiode response had little effect on the measured waveforms. The simulated fields and comparisons are shown below in Fig. S5 and S6. For both cases n=3 and n=5, they show the assumed reference signal, based on the n=5 measurement as a stand-in, and the corresponding response of the measurement assuming a flat photodiode response and also the real response of Si-based photodetector. In addition we also show the corresponding error amplitude at the lower panel.

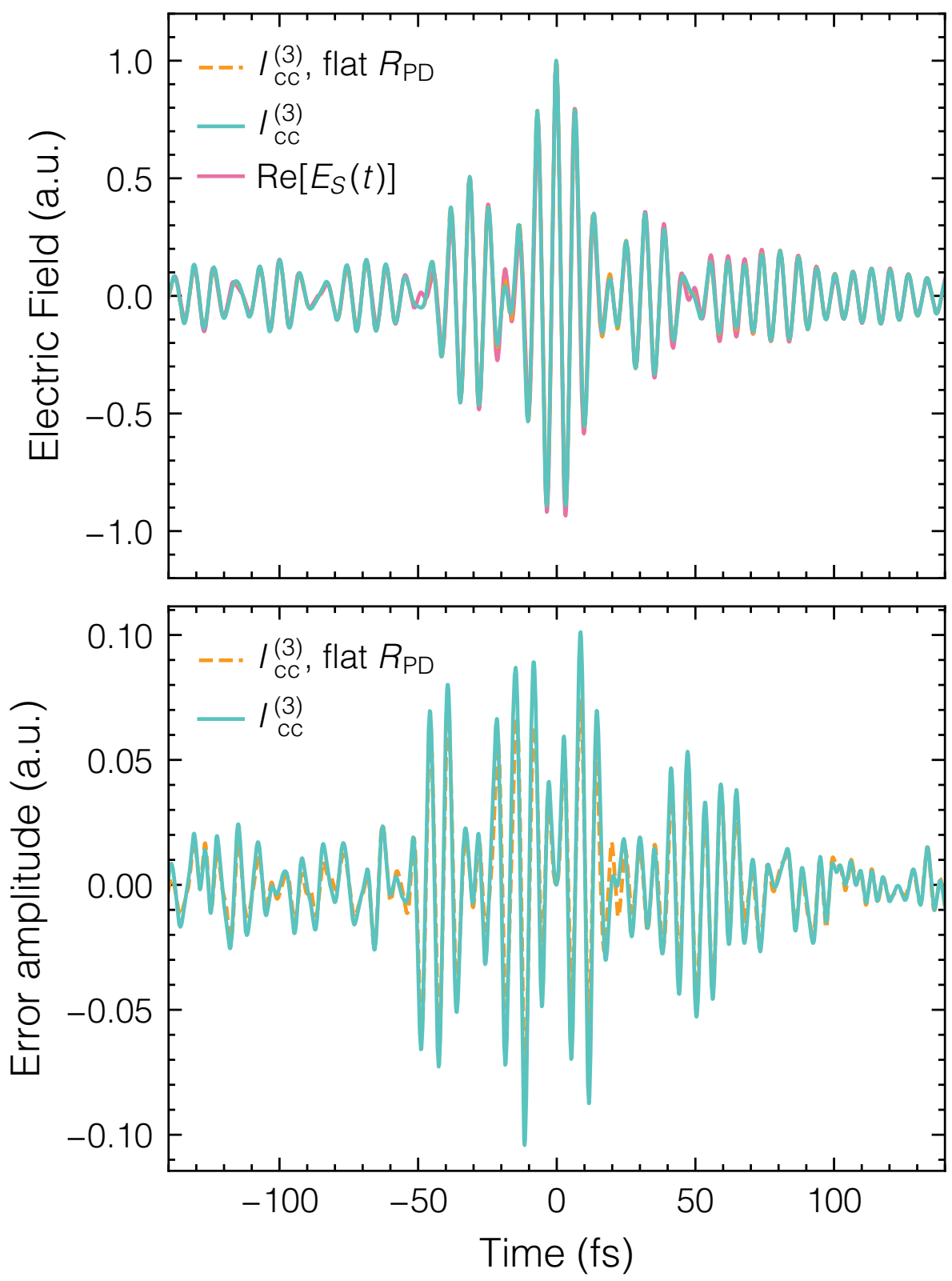


**Fig. S5. Simulated waveform comparison with and without photodiode spectral response, n=3**. Top) direct waveform comparisons. Bottom) $I_{CC}^{(3)} - \mathrm{Re}\left[E_S\right]$.

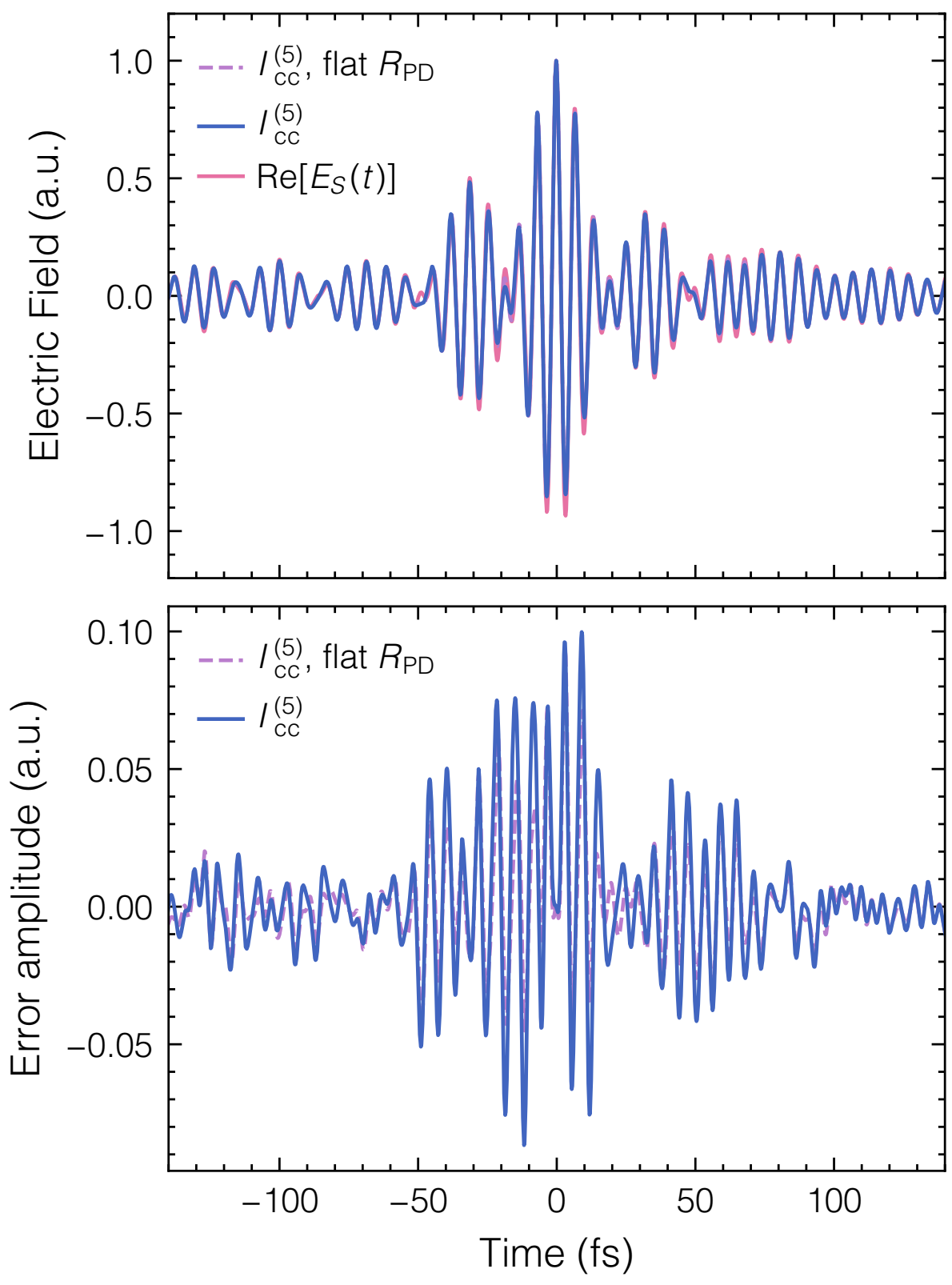


**Fig. S6. Simulated waveform comparison with and without photodiode spectral response, n=5**. Top) direct waveform comparisons. Bottom) $I_{CC}^{(5)} - \mathrm{Re}\left[E_S\right]$.



\end{document}